\documentclass[useAMS,usenatbib]{mn2e}

\include{epsf}

\def\gsim{\lower.73ex\hbox{$\sim$}\llap{\raise.4ex\hbox{$>$}}$\,$}
\def\lsim{\lower.73ex\hbox{$\sim$}\llap{\raise.4ex\hbox{$<$}}$\,$}
\def\mpc{$\,h^{-1}\,$Mpc}
\def\%{~per~cent}

\title[High-order 2MASS galaxy correlation functions]
{High-order 2MASS galaxy correlation functions: Probing the primordial density field and the linearity of galaxy bias}
\author[W.J. Frith, P.J. Outram \& T. Shanks]
{W.J. Frith\thanks{E-mail:w.j.frith@durham.ac.uk}, 
P.J. Outram \& T. Shanks\\
Dept. of Physics, Univ. of Durham, South Road, Durham DH1 3LE, UK}
\begin{document}

\date{Accepted 2005. Received 2005; in original form 2005 }

\pagerange{\pageref{firstpage}--\pageref{lastpage}} \pubyear{2005}

\maketitle

\label{firstpage}

\begin{abstract}

We use the 2 Micron All Sky Survey (2MASS) extended source catalogue to determine area-averaged angular correlation functions, $\bar{\omega}_p$, to 
high orders ($p\le$9). The main sample used contains 650$\,$745 galaxies below an extinction-corrected magnitude of $K_s$=13.5 and limited to 
$|b|>$10$^{\circ}$ and represents an order of magnitude increase in solid angle over previous samples used in such analysis ($\approx$34$\,$000 
deg$^2$). The high-order correlation functions are used to determine the projected and real space hierarchical amplitudes, 
$s_p=\bar{\omega}_p/\bar{\omega}_2^{p-1}$ and $S_p=\bar{\xi}_p/\bar{\xi}_2^{p-1}$. In contrast to recent results, for $p\le6$ these parameters are 
found to be quite constant over a wide range of scales to $r\approx$40\mpc , consistent with a Gaussian form to the primordial distribution of 
density fluctuations which has evolved under the action of gravitational instability. We test the sensitivity of our results to the presence of rare 
fluctuations in the local galaxy distribution by cutting various regions of over-density from the main sample; unlike previous analyses, we find that 
at least for $p$\lsim 4, the high-order clustering statistics are relatively robust to the removal of the largest superclusters. Since we probe well 
into the linear regime we are able to make direct comparisons with perturbation theory; we use our constraints on the $K_s$-band $S_p$ parameters in 
two ways. First, we examine their consistency with non-Gaussian initial conditions; we are able to rule out strong non-Gaussianity in the primordial 
density field, as might be seeded by topological defects such as cosmic strings or global textures, producing relations of the form 
$\bar{\xi}_p\propto\bar{\xi}_2^{p/2}$, at the $\approx$2.5$\sigma$ confidence level. Second, we investigate the way in which galaxies trace the 
underlying mass distribution. We find evidence for a non-zero quadratic contribution to the galaxy bias, parameterised by $c_2$=0.57$\pm$0.33. This 
{\em positive} result represents a significant difference from the negative values found previously; we examine a possible explanation in the light 
of recent observations which universally provide negative values for $c_2$.

\end{abstract}

\begin{keywords}
surveys - galaxies: clustering - cosmology: observations - large-scale structure of the
Universe - infrared: galaxies
\end{keywords}

\section{Introduction}

The variance of local galaxy density fluctuations has been measured to high 
accuracy over the last decade, both through the 2-point correlation function \citep[e.g.][]{zeh2,mal,haw} and its Fourier transform, 
the power spectrum \citep[e.g.][]{col3,fri2}. For a perfectly Gaussian density field, the 2-point statistic forms a complete 
description of the galaxy distribution as all higher order connected moments are zero. 

Assuming a Gaussian form for the primordial distribution of density fluctuations, perturbation theory predicts non-zero higher order correlation 
functions of the local galaxy density field; as the initial inhomogeneities grow gravitationally, the density distribution becomes asymmetric, 
developing non-zero skewness and kurtosis etc. However, it is possible to construct models of inflation, and also models which contain non-linear 
structures in the primordial density field, such that the initial conditions themselves are non-Gaussian \citep[e.g.][]{sil,wei}. Therefore, if 
non-zero high-order moments of the local galaxy density field are detected, these arise either via the gravitational collapse of 
initially Gaussian density fluctuations or from some degree of primordial non-Gaussianity.

Examining the high-order correlation functions of the local galaxy distribution therefore represents a particularly powerful approach to 
understanding the nature of the primordial density field. It is now well 
established that under the assumption of Gaussian initial conditions, these high-order moments are expected to display a hierarchical scaling 
such that the {\it p}-order cumulants of the local density field $< \delta^p >_c$ can be expressed in terms of the variance of the distribution, 
such that $<\delta^p>_c=S_p<\delta^2>^{p-1}$ \citep[e.g.][]{peb4,fry,bou2,ber,ber2,gaz3,bau2}. 

These $S_p$ coefficients, which quantify the departure from Gaussian behaviour, are insensitive to cosmic epoch or assumed cosmology at scales 
where the growth of the density field is linear or quasi-linear. Departures from the hierarchical scaling of the high-order galaxy correlation 
functions on these scales might be expected only in the case of strongly non-Gaussian initial conditions or some form of scale-dependency at large 
scales in the bias between the galaxy distribution and the underlying mass distribution. 

Several surveys have made a significant contribution to our understanding of this issue. It has long been known that high-order correlation 
functions are non-zero on small scales \citep[e.g.][]{gro,sau,bou,gaz,sza2,hoy}. More recently, \citet{cro} used volume-limited catalogues drawn from 
the 2dF Galaxy Redshift Survey (2dFGRS) to constrain $S_p$ in redshift space (for $p\le$5) to $r\approx$10\mpc. For $r$\lsim 4\mpc ~the hierarchical 
scaling parameters, $S_p$, were found to be approximately constant. However, at larger values of $r$ these coefficients increased with scale for some 
luminosities, consistent with some models of structure formation with strongly non-Gaussian initial density fields \citep{gaz5,gaz4,whi,ber3}. 
However, the results were complicated by the presence of two massive superclusters which, when removed from the analysis (corresponding to a reduction 
in the volume of $\approx$2\%), now resulted in constant scaling parameters on all scales to $r\approx$10\mpc, consistent with Gaussian initial 
conditions. It appears therefore that the 2dFGRS does not probe a large enough volume to constitute a fair sample of the local galaxy distribution 
for high-order correlation functions. 

The form of the scaling parameters on larger scales are also poorly constrained. Previously, \citet{gaz} used the APM 
galaxy survey to $b_J=20$ to constrain high-order ($p\le$9) angular correlation functions and the associated angular scaling parameters, $s_p$, 
to $\theta \approx$7$^{\circ}$ (corresponding to $r$\lsim30\mpc). Despite the fact that the APM galaxy survey covers a $\approx$4$\times$ larger 
solid angle than the 2dFGRS, a similar increase in $s_3$ on large scales was found as seen for $S_3$ observed in the 2dFGRS. At higher orders, 
further departures from the expected hierarchical scaling were also observed, although the scales probed were limited to $\theta$\lsim 3$^{\circ}$ 
and the statistics became increasingly uncertain. These features have also been measured in the smaller Edinburgh-Durham Galaxy Catalogue \citep{sza}. 
Of course, the increase in $s_3$ at large scales detected by \citet{gaz} may also be due to the same massive supercluster observed in the Southern 
2dFGRS field, as the 2dFGRS is drawn from the larger APM galaxy sample. However, it would be surprising if the volume probed by the APM galaxy survey 
still did not constitute a fair sample of the Universe. 

High-order correlation functions and the amplitude of the associated scaling parameters also provide us with a powerful probe of the way in which 
galaxies trace the underlying mass distribution. Recent measurements of the linear bias, that is the bias associated with the variance of the density 
field such that $\bar{\xi}_{2,gal} = b_1^2 \bar{\xi}_{2,DM}$, indicate that in the near infra-red $b_1>$1 \citep[e.g.][]{fri2} whereas for 
optically-selected galaxies $b_1\approx$1 \citep[e.g.][]{ver,gaz7}. Whether there exist non-linear contributions to the galaxy bias, such that the 
bias is a function of the density field, can be examined using high-order moments of the galaxy density field \citep[e.g.][]{fry2}. 

Previous high-order clustering analyses have universally found that the observed skewness etc. are significantly lower than the expected values in a 
$\Lambda$CDM cosmology. This discrepancy has been interpreted as evidence for a negative, non-linear contribution to the galaxy bias.  Most recently, 
\citet{gaz7} used volume-limited samples of the 2dFGRS to determine the redshift space 3-point correlation function and constrained $b_J$-band bias 
parameters to $b_1=0.94^{+0.13}_{-0.11}$ and a non-linear, quadratic bias of $c_2=b_2/b_1=-0.36^{+0.13}_{-0.09}$, although as with previous 2dFGRS 
analyses these results are significantly affected by the presence of two massive superclusters. Independently, \citet{pan} determined the monopole 
contribution to the 2dFGRS 3-point correlation function and determined a similar $b_J$-band linear bias to \citet{gaz7} and a negative $c_2$ parameter 
as well, although considerably smaller and at a reduced significance to the other 2dFGRS constraint. Computing the bispectrum for the PSCz catalogue 
(which is selected from the 60$\mu$m IRAS galaxy sample), \citet{fel2} also constrain infra-red bias parameters of $b_1=0.83\pm0.13$ and 
$c_2=-0.50\pm0.48$.

The 2 Micron All Sky Survey (2MASS) has recently been completed and provides $K_s$, $H$ and $J$-band photometry for 1.6$\times$10$^6$ extended
sources over the entire sky to $K_s$\gsim13.5. 2MASS is the largest existing all-sky galaxy survey and therefore represents a uniquely 
powerful probe of the local galaxy density field at large scales; the solid angle of the 2MASS $|b|\ge$10$^{\circ}$ sample used in this 
paper represents an order of magnitude increase over the APM galaxy survey, meaning that clustering statistics determined from the 2MASS galaxy 
sample will suffer less from projection effects while probing a comparable volume. The 2MASS survey also represents an order of magnitude increase in 
volume over the largest volume-limited 2dFGRS sample; it is therefore possible to probe much larger scales than the 2dFGRS although with the added 
complication of projection effects. A further advantage of 2MASS over previous datasets is that the photometry is extremely accurate with high 
completeness for $|b|\ge$10$^{\circ}$; correcting for the variable completeness over the 2dFGRS survey area for instance, complicates the analysis 
and increases the uncertainty from possible systematic effects. The main drawback to high-order clustering analysis of the 2MASS data (as with 
the APM galaxy survey) is the lack of availiable three-dimensional information; the clustering signal from a particular scale in real space is 
smeared over a range of angular scales. For this reason detailed features in the real space correlation function, such as the shoulder at 
$r\approx$10\mpc ~\citep{bau4,haw,zeh2}, may not be detected by the 2MASS projected correlation function.

In this work, we aim to determine the high-order angular correlation functions and the associated scaling parameters of the local galaxy density 
field to high precision and large scales, using the final 2MASS extended source catalogue. In section 2, the details of the method for 
estimating high-order correlation functions are given. The galaxy sample used and the error analysis are described in section 3. We present the 
$p$-point angular correlation functions (for $p\le$9) and the hierarchical scaling relations in section 4. We also examine possible systematic 
effects arising from extreme fluctuations in the observed galaxy density field. The implications for the form of primordial density 
fluctuations and non-linear galaxy bias are discussed in section 5. The conclusions follow in section 6.

\section{Method of Estimation}

\subsection{The {\it p}-point correlation function}

The {\it p}-point galaxy correlation function estimates the joint probability that {\it p} galaxies are 
separated by a certain scale and can be defined through considering fluctuations in the 
galaxy density field. The connected or reduced part of this statistic corresponds to the contribution 
to this probability which does not include any conditional probability on lower orders:

\begin{equation}
\omega_p(\theta_1,...,\theta_p) \equiv < \delta_1,...,\delta_p >_c ,
\label{equation:wp1}
\end{equation}

\noindent where $\delta$ denotes the density fluctuation; for $p\le$3, the unreduced and reduced correlation functions 
are the same. In this paper, we work with the reduced {\it p}-point correlation function only.

The 2-point angular galaxy correlation function, $\omega_2(\theta)$, is given in terms of the probability of finding two galaxies 
in area elements $d\Omega_1$ and $d\Omega_2$ separated by angle $\theta$:

\begin{equation}
dP_2 = {\mathcal{N}}^2~ [1+\omega_2(\theta)] ~d\Omega_1 d\Omega_2 ,
\label{equation:w2}
\end{equation}

\noindent where ${\mathcal{N}}$ is the mean number of galaxies per unit solid angle \citep[e.g.][]{gro,peb4}. Similarly, the 3-point function, 
$\omega_3(\theta)$, is defined by the joint probability of finding galaxies in each of three area elements:

\[ dP_3 = {\mathcal{N}}^3~ [1+\omega_2(\theta_{12})+\omega_2(\theta_{23})+\omega_2(\theta_{13}) \]

\begin{equation}
~~~~~~~ +\omega_3(\theta_{12},\theta_{23},\theta_{31})] ~d\Omega_1 d\Omega_2 d\Omega_3
\label{equation:w3}
\end{equation}

\noindent The first terms in equations~\ref{equation:w2} and \ref{equation:w3} correspond to the contributions from galaxy pairs or triplets 
respectively which are accidentally seen as close together in projection but are at very different radial distances. Similarly, the following three 
terms in equation~\ref{equation:w3} account for the contribution from one correlated pair and a third uncorrelated galaxy which forms a triplet by 
chance line-of-sight clustering. This leaves the final term which defines the contribution from the real clustering of triplets.

A simple way in which to estimate the high-order correlation functions is through the area-averaged correlation function, $\bar{\omega}_p$, 
defined as:

\begin{equation}
\bar{\omega}_p (\theta) = \frac{1}{\Omega^p} \int_{\Omega} \omega_p (\theta_1,...,\theta_p) ~d\Omega_1,...,d\Omega_p ,
\label{equation:wbar}
\end{equation}

\noindent where $\Omega$ is the solid angle of the cone defined by its angular radius $\theta$. The area-averaged correlation function, 
$\bar{\omega}_p$, is estimated by considering the central moments of the angular counts:

\begin{equation}
m_p(\theta) = < (N-\bar{N})^p > = \sum_{N=0}^{N=\infty} (N-\bar{N})^p P_N(\theta) ,
\label{equation:moments}
\end{equation}

\noindent where $P_N(\theta)$ denotes the count probability distribution function and is calculated by placing circular cells of angular 
radius $\theta$ over the survey area and determing the number of cells containing exactly {\it N} galaxies:

\begin{equation}
P_N (\theta) = \frac{N_N}{N_C} ,
\label{equation:cpdf}
\end{equation}

\noindent where {\it N$_N$} and {\it N$_C$} denote the number of cells containing {\it N} galaxies and the total number of cells respectively.
$\bar{N}$ in equation~\ref{equation:moments} is the mean number of galaxies in a cell of angular radius $\theta$ and may be determined directly 
from the count probability distribution function:

\begin{equation}
\bar{N} = \sum_{N=0}^{N=\infty} N P_N(\theta)
\label{equation:nbar}
\end{equation}

The moments of the count probability distribution function determined via equation~\ref{equation:moments} yield the unreduced correlation function 
through the relation $m_p = < \delta^p > \bar{N}^p$. In order to obtain the reduced correlation function, the connected moments $\mu_p$ are 
determined:

\[ \mu_2 = m_2 \]
\[ \mu_3 = m_3 \]
\begin{equation}
\mu_4 = m_4 - 3m_2^2 ,
\label{equation:mu}
\end{equation}

\noindent (see \citet{gaz} for higher order relations). In addition we apply a shot noise correction \citep{gaz,bau3} such that:

\[ k_2 = \mu_2 - \bar{N} \]
\[ k_3 = \mu_3 - 3k_2 - \bar{N} \]
\begin{equation}
k_4 = \mu_4 - 7k_2 - 6k_3 - \bar{N}
\label{equation:k}
\end{equation}

\noindent The area-averaged, reduced angular correlation function is then determined from the relation 
$\bar{\omega}_p = < \delta^p >_c = k_p/\bar{N}^p$.

\subsection{Hierarchical scaling}

In perturbation theory, the density field, evolved by gravity from an initially Gaussian distribution, 
leads to a hierarchical clustering pattern, such that all high-order correlations can be expressed in 
terms of the 2-point correlation function:

\begin{equation}
\bar{\xi}_{p,\mathrm{DM}} = S_{p,\mathrm{DM}}~ \bar{\xi}_{2,\mathrm{DM}}^{p-1} ,
\label{equation:3dhier}
\end{equation}

\noindent where $\bar{\xi}_{p,DM}$ is the volume-averaged {\it p}-point dark matter correlation function. Importantly, this relation is preserved for 
the galaxy density field such that $\bar{\xi}_{p,gal} = S_{p,gal}~ \bar{\xi}_{2,gal}^{p-1}$ \citep{fry2}. The relation between the dark matter and 
galaxy density fields may be expressed through a Taylor expansion of the dark matter density contrast:

\begin{equation}
\delta_{\mathrm{gal}} = \sum_{p=0}^{\infty} \frac{b_p}{p!}(\delta_{\mathrm{DM}})^p
\label{equation:taylor}
\end{equation}

\noindent For the skewness, it can be shown that \citep{fry2}:

\begin{equation}
S_{3,\mathrm{gal}} = \frac{1}{b_1}(S_{3,\mathrm{DM}}+3c_2) ,
\label{equation:skew}
\end{equation}

\noindent where $b_1$ is the linear bias such that $\bar{\xi}_{2,gal} = b_1^2 \bar{\xi}_{2,DM}$ on scales where the variance of the density field is 
small, and $c_2=b_2/b_1$ quantifies the second-order contribution to the galaxy bias.

Using perturbation theory, it is possible to determine precise quantitative predictions for the $S_{p,DM}$ parameters. Assuming a power law 
form for the three-dimensional power spectrum of density fluctuations, $P(k)\propto k^n$, and a top-hat window function, the skewness of the matter 
distribution, $S_{3,DM}$ may be determined \citep{jus}:

\begin{equation}
S_{3,\mathrm{DM}}=\frac{34}{7}-($n$+3)
\label{equation:2dpt}
\end{equation}

\noindent We later use these expressions to determine constraints on non-linear bias through comparisons with $S_{3,gal}$ in the linear and 
quasi-linear regimes. We assume a power spectrum slope of $n$=-2 \citep{per,col3}. The uncertainties on $n$ are small ($\Delta n<$0.1) compared to 
the sampling errors in the measurement of $S_{3,gal}$; we therefore assume the concordance value cited above to generate predictions for $S_{3,DM}$ 
and neglect small uncertainties in this parameter.

\subsection{Transformation to three dimensions}

For a projected galaxy distribution, a hierarchical scaling relation may also be defined in terms of the area-averaged {\it p}-point galaxy 
correlation functions and angular scaling coefficients, $s_p$:

\begin{equation}
\bar{\omega}_p = s_p~ \bar{\omega}_2^{p-1}
\label{equation:2dhier}
\end{equation}

\noindent We wish to transform these angular scaling parameters, $s_p$, to the three-dimensional coefficients, $S_p$, in order to make comparisons 
with perturbation theory and constrain non-linear galaxy bias. Following the method of \citet{gaz}, we transform to three dimensions via the 
relation:

\begin{equation}
S_p(\bar{r}) \approx \frac{s_p(\theta)B_p(\gamma)}{r_p(\gamma)C_p(\gamma)} ,
\label{equation:3dpt}
\end{equation}

\noindent where $\bar{r}=\theta\mathcal{D}$ is the mean scale probed at an angular scale $\theta$ for a survey of median depth $\mathcal{D}$, 
$\gamma$ is the slope of the 2-point real space correlation function, and $B_p$ and $C_p$ are related to the number of different configurations of 
the three-dimensional and angular hierarchical tree graphs respectively \citep{gaz}. Here we use the form of the real space correlation function 
determined from the 2dFGRS \citep{haw}. The $r_p$ factor is related to the selection function $\Psi$:

\begin{equation}
r_p = \frac{I_1^{p-2}I_p}{I_2^{p-1}}
\label{equation:rp}
\end{equation}

\noindent where 

\begin{equation}
I_j = \int_{0}^{\infty} \Psi^j x^{(3-\gamma)(j-1)}(1+z)^{(3+\epsilon-\gamma)(1-j)} F(x) x^2 dx 
\label{equation:ij}
\end{equation}

\noindent where $\epsilon$ describes the evolution of clustering with redshift \citep{gro} and is taken to be $\epsilon$=0 in good agreement with 
recent observational and theoretical considerations \citep{ham,pea,car,wil}. Here, $x$ denotes the comoving distance and $F(x)$ a correction for 
curvature such that $F(x)=[1-(H_0x/c)^2(\Omega_m-1)]^{1/2}$. In this work we use the concordance value of $\Omega_m$=0.3. For the selection function 
we use a parameterised form for the $n(z)$ such that: 

\begin{equation}
n(z)=\frac{3z^2}{2(\bar{z}/1.412)^3} \exp \left(-\left(\frac{1.412z}{\bar{z}}\right)^{3/2}\right)
\label{equation:sel}
\end{equation}

\noindent \citep{bau,mal} where $\bar{z}$ is determined from the 2MASS-2dFGRS matched sample described in \citet{fri4}. For reference 
$\bar{z}$=0.074 for $K_s<$13.5. In this case, the normalisation of the selection function factors out (see equations~\ref{equation:rp} and 
\ref{equation:ij}).

The transformation described in equation~\ref{equation:3dpt} is robust to reasonable changes in the selection function and 
choice of cosmological parameters \citep{gaz}. However, this relation becomes uncertain on large angular scales, $\theta>$2$^{\circ}$ \citep{gaz}, 
due to the fact that the power law form to the 2-point correlation function and the value of $\gamma$ are not well 
constrained on large scales. 

\section{Analysis of the 2MASS data}

\subsection{The 2MASS Extended Source Catalogue}

We select objects from the 2MASS final release extended source catalogue \citep{jar2} above a galactic latitude of $|b|=10^{\circ}$ in order to remove 
regions of high extinction and stellar contamination (see Fig.~\ref{fig:sky}) and below an extinction-corrected magnitude limit of $K_s$=13.5 using 
the dust maps of \citet{sch}. The 2MASS dataset removes or flags sources identified as artefacts such as diffraction spikes and meteor streaks 
\citep{jar}; we use the 2MASS $cc\_flag$ to remove such objects. We also employ a colour cut ($J-K_s<$0.7 and $J-K_s>$1.4) below $K_s$=12 in order to 
remove a small number of objects identified as non-extragalactic extended sources \citep{mal2,mal}. The subsequent sample of 650$\,$745 galaxies 
probes to a median depth of $\mathcal{D}\approx$220\mpc~and covers 83\% of the entire sky ($\approx$34$\,$000 deg$^2$).

We estimate the total $K_s$-band flux using the magnitude estimator described in \citet{fri4,fri2}, based on the method of \citet{cole}. To recap, we 
use the deeper and more accurate 2MASS $J$-band extrapolated magnitudes and colour-correct to the $K_s$-band using the $J$ and $K_s$-band fiducial 
elliptical Kron magnitudes; this provides excellent agreement with the independent $K$-band photometry of \citet{lov} and also the 2MASS-selected 6dF 
Galaxy Survey total magnitude estimator \citep{jon}.

In order to calculate the high-order correlation functions of this 2MASS sample, we determine the count probability distribution function 
detailed in equation~\ref{equation:cpdf} by randomly placing $N_C$=10$^6$ cell centres over the survey area. Each cell is then 
allowed to grow and the number of galaxies as a function of the angular radius is recorded. The size of the cells is limited by the galactic 
latitude limit of $|b|=10^{\circ}$ in the 2MASS sample. We replace cells which are lost as they encroach the boundary of the 
galaxy sample such that $N_C$=10$^6$, independent of the angular scale probed. We use cell radii in the range 0$^{\circ}$.01 to 25$^{\circ}$.1, 
equivalent to a range in the mean scale probed of 0.04\mpc~to 104\mpc .

\subsection{Error estimation}

\begin{figure*}
\vspace*{1cm}
\begin{center}
\centerline{\epsfxsize = 7in
\epsfbox{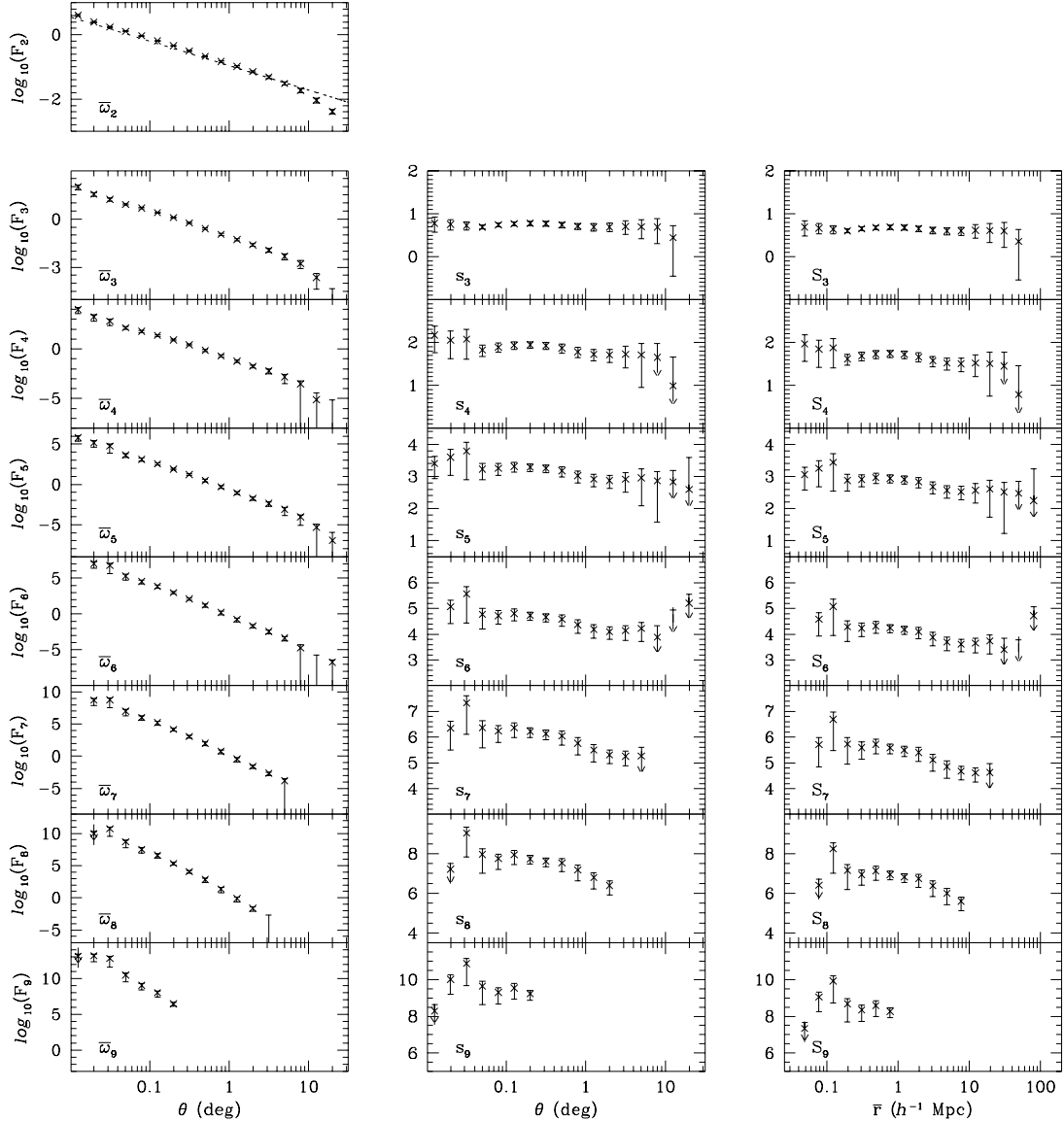}}
\vspace*{-0.5cm}
\caption{Moments of the 2MASS galaxy density field for the full $|b|\ge10^{\circ}$ $K_s<$13.5 sample (all on logarithmic scales).
Each row corresponds to a different moment of the galaxy density field. Since in each column we plot a different statistic ($\omega_p$, $s_p$ or 
$S_p$) we denote the $x$-axis label as F$_p$. In each case the errors are taken from bootstrap estimates described in
section 3.2. Datapoints with extremely large errorbars are omitted for clarity. The columns are set out as follows: \newline
{\bf First column:} Area-averaged correlation functions for 650$\,$745 $K_s<$13.5 2MASS galaxies. In the $\bar{\omega}_2$ panel, the best fit
result at small scales from \citet{mal} is indicated by the dotted line. \newline
{\bf Second column:} The angular scaling parameters ($s_p$) determined via equation~\ref{equation:2dhier} for the full sample. \newline
{\bf Third column:} The real space scaling parameters ($S_p$) determined via equation~\ref{equation:3dpt} for the full sample. 
}
\label{fig:wp}
\end{center}
\vspace*{1cm}
\end{figure*}

\begin{figure*}
\begin{center}
\centerline{\epsfxsize = 7in
\epsfbox{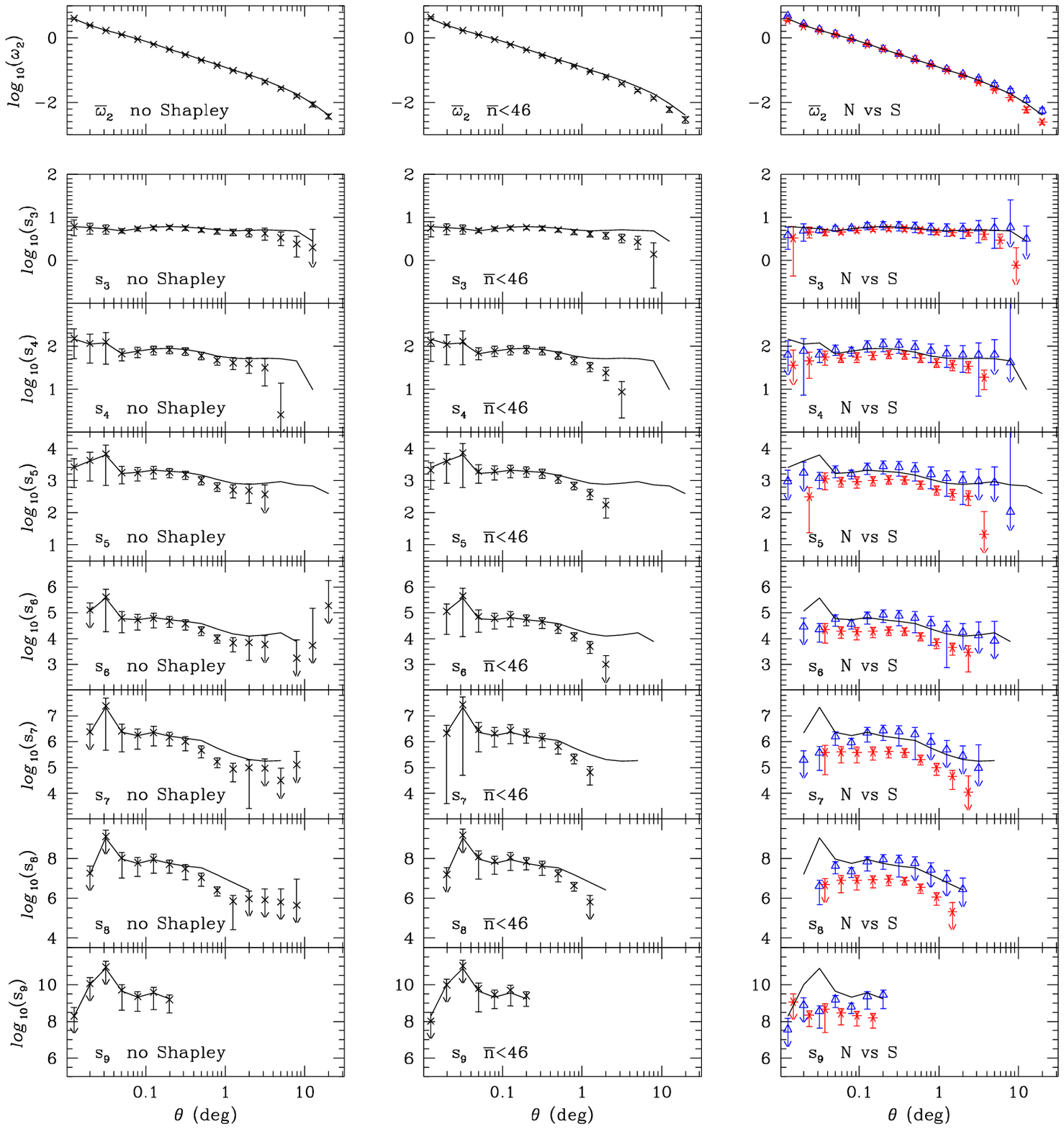}}
\caption{Moments of the 2MASS galaxy density field making various cuts to the full $|b|\ge10^{\circ}$ $K_s<$13.5 sample. As in Fig.~\ref{fig:wp} each
row corresponds to a different moment and the errors are determined via bootstrap estimates as described in section 3.2. For $p$=2 (the top row) we
show the area-averaged correlation function; for higher order moments we display the angular scaling parameters, $s_p$. In each case we indicate the
corresponding results from the full sample shown in Fig.~\ref{fig:wp} by a solid line. Datapoints with extremely large errorbars are omitted for 
clarity. The columns are set out as follows: \newline
{\bf First column:} We show the 2-point function and the higher-order angular scaling parameters having removed a region of radius 6$^{\circ}$
centred on the largest over-density in the sample, the Shapley supercluster (see section 4.2). This corresponds to a removal of 1.1\% of the galaxies
and 0.33\% of the solid angle of the full sample. \newline
{\bf Second column:} We apply a more stringent cut such that areas of radius 6$^{\circ}$ centred on the ten most over-dense pixels in
Fig.~\ref{fig:sky} are removed (see table~\ref{table:clusters}) corresponding to a removal of 6.3\% of the galaxies and 2.6\% of the solid angle of
the full sample. \newline
{\bf Third column:} Here, we split the full sample into north (triangles) and south (stars). For clarity, we have displaced the southern angular
scaling parameter datapoints to the right. }
\label{fig:wp2}
\end{center}
\end{figure*}

The statistical uncertainty associated with the correlation function and angular scaling parameters in this work are determined using bootstrap 
estimates. The full 2MASS sample is split into 20 equal area regions of $\approx$1700 deg$^2$; 20 of these sub-areas are selected at random (with 
repeats) and the associated clustering statistic is determined. We repeat this 1000 times and determine the standard deviation; these are indicated 
by the errorbars in Figs.~\ref{fig:wp} and \ref{fig:wp2}. The size of these errors is not significantly altered if we vary the number of realisations 
or sub-areas within reasonable limits. 

Once we have determined the angular correlation functions and scaling parameters and the associated errors in this way, we wish to carry out 
comparisons with predictions from perturbation theory. Since correlation function estimates determined at different cell radii are highly 
correlated being integral quantities, it is necessary to account for the covariance between each datapoint when performing fits to the data. We 
determine the covariance matrix ($C_{ij}$) using the bootstrap method above. We then determine $\chi^2$:

\begin{equation}
\chi^2 = \sum_{i} \sum_{j} \Delta_i C_{ij}^{-1} \Delta_j ,
\label{equation:chi}
\end{equation}

\noindent where $\Delta_i = s_p^{obs}(i) - s_p^{mod}(i)$ for example. We compute the inverse covariance matrix, $C_{ij}^{-1}$, using the 
Numerical Recipes Singular Value Decomposition algorithm. As noted by \citet{cro}, previous constraints on the high-order scaling parameters 
(except for the 2dFGRS results) ignored the correlations between different bins leading to unrealistically small errors in the fitted values.

\section{Results}

\subsection{Area-averaged correlation functions}

In the first column of Fig.~\ref{fig:wp} we plot the area-averaged correlation functions for $p\le$9 determined for the full $|b|\ge$10$^{\circ}$ 
2MASS galaxy sample described in section 3.1. In each case the errors are determined via the bootstrap technique described in section 3.2. For 
reference we compare our result for $\bar{\omega}_2$ with the best fit to the 2MASS $K_s<$13.5 2-point correlation function of \citet{mal} 
($A$=0.10$\pm$0.01, 1-$\gamma$=-0.79$\pm$0.02, determined on scales of $\theta\le$2$^{\circ}.5$ where $\omega_2=A\theta^{1-\gamma}$). 

The $p$-point correlation functions are consistent with power laws over several orders of magnitude in angular scale. In particular, we note that the 
area-averaged $\bar{\omega}_2$ determined in this work is in good agreement with the directly-determined result of \citet{mal} at small scales 
(we find $A=0.13\pm0.01$, 1-$\gamma$=-0.77$\pm$0.04 for $\theta\le$2$^{\circ}.5$). This agreement is robust to changes in the magnitude 
estimator and galactic latitude cut; \citet{mal} use the $K_s$-band elliptical Kron magnitude estimator and a $|b|\ge20^{\circ}$ cut.

\begin{table}
\centering
\begin{tabular}{|p{0.3cm}|p{2.7cm}|p{2.7cm}|} \hline
$p$     & ~~~~~~$s_{p}$      &       ~~~~~~$S_{p}$                \\
\hline
3               & 5.28~$\pm$~0.45                       & 4.29~$\pm$~0.39                               \\
4               & 57.7~$\pm$~9.2                        & 36.0~$\pm$~5.7                                \\
5               & 1510~$\pm$~507                        & 678~$\pm$~228                                 \\
6               & (3.68~$\pm$~2.08)$\times$10$^4$       & (1.20~$\pm$~0.68)$\times$10$^4$               \\
7               & (9.74~$\pm$~8.72)$\times$10$^5$       & (2.25~$\pm$~2.01)$\times$10$^5$               \\
8               & (2.64~$\pm$~3.69)$\times$10$^7$       & (4.19~$\pm$~5.86)$\times$10$^6$               \\
9               & (2.03~$\pm$~1.56)$\times$10$^9$       & (2.26~$\pm$~1.74)$\times$10$^8$               \\
\hline
\end{tabular}
\caption{Small scale fits to the high-order scaling parameters (assuming constant values) determined for the full 2MASS $|b|\ge10^{\circ}$, $K_s<$13.5
sample (see Fig.~\ref{fig:wp}). In the second column the best fit angular scaling parameters (for 3$\le p\le$9) are shown, determined from $\chi^2$
fits in the range 0$^{\circ}.04<\theta<$1$^{\circ}.0$ for $p\le$8 and 0$^{\circ}.04<\theta<$0$^{\circ}.25$ for $p$=9. Similarly, in the third column
we show the best fit real space scaling parameters, $S_p$, fitted over an equivalent range of scales, 0.15$<\bar{r}<$4.0\mpc ~for $p\le$8 and
0.15$<\bar{r}<$1.0\mpc ~for $p$=9. The errors in each case take into account the covariance matrix determined from the bootstrap estimates
described in section 3.2. }
\label{table:sptab}
\end{table}

\begin{table}
\centering
\begin{tabular}{|p{0.3cm}|p{2.7cm}|p{2.7cm}|} \hline
$p$     & ~~~~~~$s_{p}$      &       ~~~~~~$S_{p}$                \\
\hline
3               & 4.91~$\pm$~0.60                       & 4.00~$\pm$~0.49                               \\
4               & 54.2~$\pm$~11.1                       & 33.8~$\pm$~6.9                                \\
5               & 740~$\pm$~240                         & 332~$\pm$~108                                 \\
6               & (1.01~$\pm$~0.67)$\times$10$^4$       & (3.30~$\pm$~2.19)$\times$10$^3$               \\
\hline
\end{tabular}
\caption{Large scale fits to the high-order scaling parameters (assuming constant values) determined for the full 2MASS $|b|\ge10^{\circ}$, $K_s<$13.5
sample (see Fig.~\ref{fig:wp}). In the second column the best fit angular scaling parameters (for 3$\le p\le$6) are shown, determined from $\chi^2$
fits in the range 1$^{\circ}.0<\theta<$10$^{\circ}$. Similarly, in the third column we show the best fit real space scaling parameters, $S_p$, fitted
over an equivalent range of scales, 4.0$<\bar{r}<$40\mpc . The errors are determined as in Table~\ref{table:sptab}.}
\label{table:sptab2}
\end{table}

In order to determine whether these results are consistent with the hierarchical scaling described in section 2.2 we 
compute the angular scaling parameters, $s_p$ (see equation~\ref{equation:2dhier}); these are shown in the second column of Fig.~\ref{fig:wp}. 
The angular scaling parameters are transformed into the real space $S_p$ parameters in the third column of Fig.~\ref{fig:wp} (see 
equation~\ref{equation:3dpt}). The $s_p$ and $S_p$ coefficients are consistent with constant values to large scales ($\theta$\lsim 20$^{\circ}$, 
$\bar{r}$\lsim 100\mpc), although there may be a decrease in amplitude between the non-linear and quasi-linear regime  
($\theta\approx$1$^{\circ}$, $r\approx$4\mpc ). One caveat to this is that while these coefficients are consistent with a slope of zero, the 
constraints become increasingly weak at higher orders such that in the range 1$^{\circ}.0<\theta<$10$^{\circ}$ the slopes are constrained to  
$\gamma(s_3)=0.01^{+0.34}_{-0.43}$, $\gamma(s_4)=0.02^{+0.45}_{-0.84}$, $\gamma(s_5)=-0.01^{+0.63}_{-0.94}$ and $\gamma(s_6)=-0.39^{+0.88}_{-0.91}$ 
at 1$\sigma$ confidence (where $s_p\propto \theta^{\gamma}$, marginalising over the normalisation); we investigate the constraint on the slope of 
$S_3$ and the level to which this can reject primordial non-Gaussianity in section 5.1.
   
Since we probe well into the linear and quasi-linear regimes we are able to make comparisons with predictions from 
perturbation theory. We perform $\chi^2$ fits to these scaling parameters considering the covariance in the datapoints (see 
equation~\ref{equation:chi}). We consider small and large scales separately due to the fact that, despite the consistency of the scaling 
parameters over all scales considered, perturbation theory is not expected to be valid on small scales \citep{ber3}; approximately, the scales used 
represent fits in the non-linear and quasi-linear or linear regimes. Also, it is important to remember that the conversion from angular to real space 
scaling parameters becomes increasingly uncertain at large angular scales ($\theta>$2$^{\circ}$). The best fit scaling parameters are shown in 
Table~\ref{table:sptab} for small scales and Table~\ref{table:sptab2} for large scales.

\begin{table}
\centering
\begin{tabular}{|p{0.2cm}|p{0.8cm}|p{0.8cm}|p{1.5cm}|p{3.0cm}|}
\hline
   & $l$ ($^{\circ}$) & $b$ ($^{\circ}$) & $\bar{n}$ (deg$^{-2}$) & Cluster \\
\hline
1  & 312.2      & ~30.0 & 69.1  & Shapley supercluster       \\
2  & 266.5      & -51.3 & 54.4  & Horologium-Reticulum  \\
3  & 45.0       & ~57.4 & 52.4  & Centaurus             \\
4  & 10.4       & ~51.3 & 52.2  & Centaurus             \\
5  & 303.8      & ~32.8 & 49.4  & Shapley supercluster       \\
6  & 266.8      & -48.1 & 47.5  & Horologium-Reticulum  \\
7  & 219.4      & -32.8 & 47.5  & NGC 1600 Group        \\
8  & 343.1      & -32.8 & 46.8  & Pavo-Indus wall       \\
9  & 312.2      & ~35.7 & 46.2  & Shapley supercluster       \\
10 & 9.6        & ~48.1 & 46.1  & Centaurus             \\
\hline
\end{tabular}
\caption{The positions and galaxy densities of the ten most over-dense pixels of the smoothed $K_s<13.5$ 2MASS galaxy distribution shown in
Fig.~\ref{fig:sky}. We also note the cluster with which each pixel is associated.}
\label{table:clusters}
\end{table}

\subsection{Fair sample issues}

It was noted in section 1, that a considerable problem in previous high-order clustering analyses is the presence of extreme fluctuations in 
the galaxy samples which have a significant effect on the observed scaling parameters. The APM and 2dF Galaxy 
Redshift Surveys observe rising $S_p$ values for $r$\gsim4\mpc ~for example, consistent with some models of structure formation with strongly 
non-Gaussian inital consitions \citep{gaz4,gaz5,ber3}; with the 2dFGRS at least the clustering signal is significantly altered when two 
superclusters are removed from the sample (corresponding to a reduction in the volume of $\approx$2\%) yielding constant $S_p$ parameters for 
$p\le5$ to scales of $r\approx$10\mpc ~\citep{bau2,cro}. 

\begin{figure*}
\begin{center}
\centerline{\epsfxsize = 7in
\epsfbox{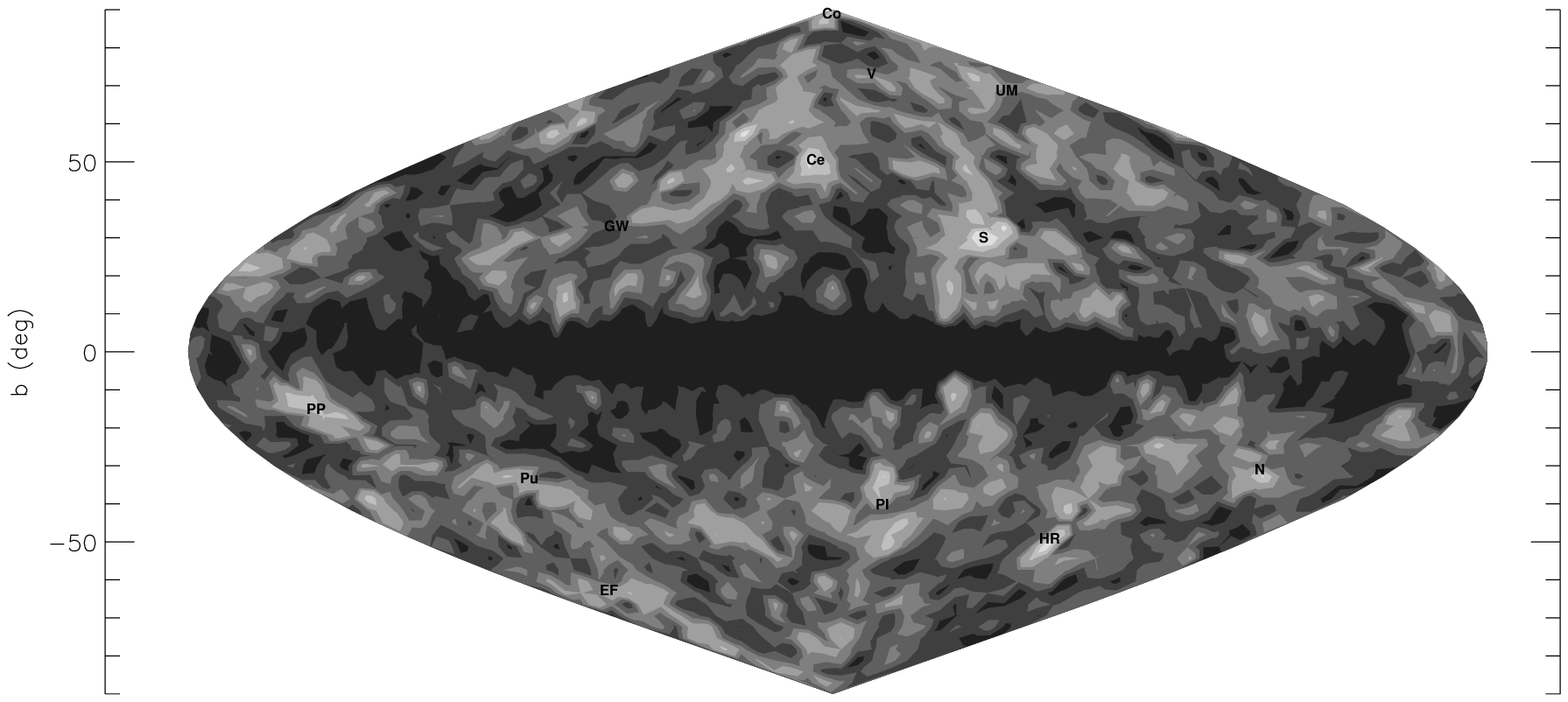}}
\vspace*{-1.8cm}
\centerline{\epsfxsize = 2.5in
\epsfbox{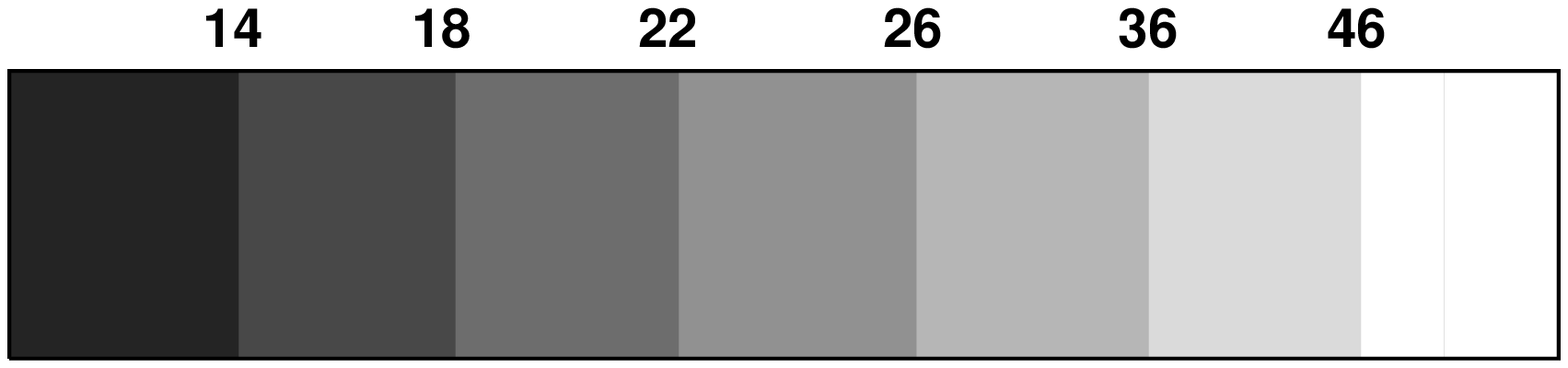}}
\centerline{\epsfxsize = 7in
\epsfbox{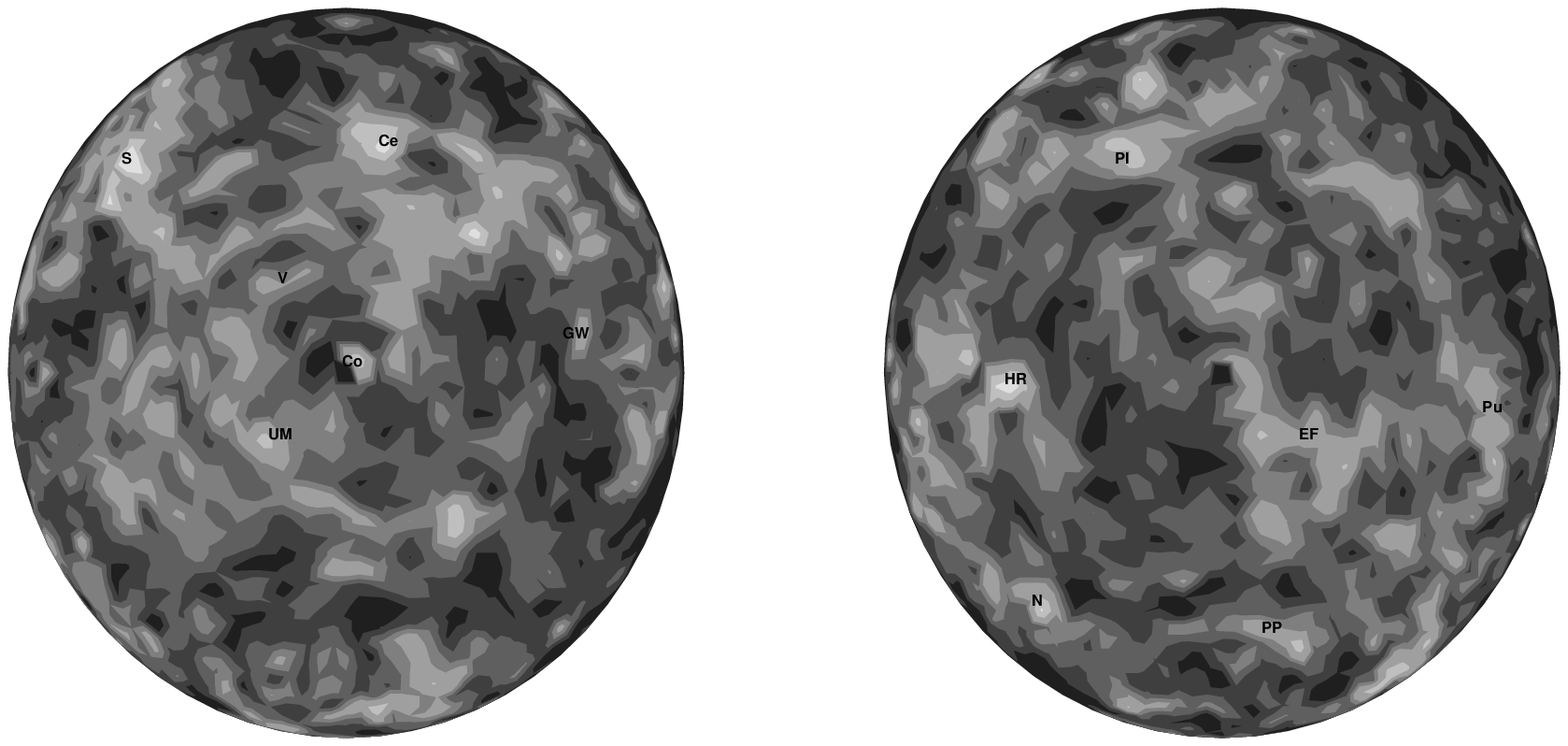}}
\caption{The number density (in deg$^{-2}$) of 2MASS $K_s<$13.5 galaxies binned in 13.5 deg$^2$ pixels; under-dense regions are
indicated by the dark filled contours; areas of over-density by lighter filled contours. For reference, the mean $|b|\ge10^{\circ}$ number density
is 19.1 deg$^{-2}$. In the upper plot we show the entire sky in Aitoff projection; in the lower panels we show the $b\ge$0$^{\circ}$ (left-hand plot)
and $b\le$0$^{\circ}$ (right-hand plot) hemispheres, such that the galactic poles are positioned in the centres of each. Prominent clusters are
indicated as follows: Co - Coma cluster, S - Shapley supercluster, V - Virgo supercluster, UM - Ursa Major cloud, GW - Great Wall, Ce - Centaurus,
HR - Horologium-Reticulum, PP - Perseus-Pisces chain, PI - Pavo-Indus wall, N - NGC 1600 Group, Pu - Puppis, EF - Eridanus-Fornax.}
\label{fig:sky}
\end{center}
\end{figure*}

We therefore wish to examine the level of such systematic effects by removing large over-densities 
from the 2MASS galaxy sample and examining the effect on the observed scaling parameters.
In Fig.~\ref{fig:sky}, we show the pixelated 2MASS $K_s<$13.5 galaxy density distribution smoothed on $\approx$7$^{\circ}$ scales (each pixel has 
a solid angle of 13.5 deg$^2$); clusters are indicated by lighter filled contours and regions of under-density by darker shades. In the upper panel 
we show the entire $K_s<$13.5 2MASS sample in Aitoff projection; in the lower panels the galaxy density field for each galactic hemisphere is plotted 
in projection separately. We also list the ten most over-dense pixels (corresponding to a limit in the number density of $\bar{n}>$46 deg$^{-2}$) in 
table~\ref{table:clusters}.

In Fig.~\ref{fig:wp2}, we plot $\bar{\omega}_2$ and the high-order angular scaling parameters with various cuts to the full galaxy sample used 
in Fig.~\ref{fig:wp}. In each case, the corresponding result for the full sample is indicated by a solid line. First, we omit the largest supercluster 
only; in the first column, we have removed a circular region with an angular radius of 6$^{\circ}$ centred on the most over-dense pixel in 
Fig.~\ref{fig:sky} sampling the Shapley supercluster. This corresponds to a removal of 1.1\% of the galaxies and 0.33\% of the solid angle of the 
full $|b|\ge$10$^{\circ}$ sample. The effect of this removal on the 2-point correlation function is insignificant. The form of $s_3$ remains 
consistent with the result for the full sample, although the best fit slope at large scales (1$^{\circ}.0<\theta<$10$^{\circ}$) changes to 
$\gamma=-0.27^{+0.25}_{-0.29}$ (compared to $\gamma=0.01^{+0.34}_{-0.43}$ for the full sample). At higher orders, there is also no significant 
effect although the statistical uncertainty increases at large scales (datapoints with extremely large errorbars are omitted for clarity). 

In the second column in Fig.~\ref{fig:wp2} we omit all galaxies within 6$^{\circ}$ of the ten most over-dense pixel centres (see 
table~\ref{table:clusters}). This corresponds to a removal of 6.3\% of the galaxies and 2.6\% of the solid angle of the full sample. There is a 
small effect on the 2-point correlation function at large scales ($\theta$\gsim 10$^{\circ}$). The effect on the higher order angular scaling 
parameters also becomes more significant. However the effect on $s_3$ at least is limited; the best fit slope at large scales 
(1$^{\circ}.0<\theta<$10$^{\circ}$) is $\gamma=-0.33^{+0.26}_{-0.29}$.

It has previously been observed that the form of the galaxy distribution in the northern and southern galactic caps are significantly different in 
many respects; \citet{mal} detected a difference in the $b\ge$20$^{\circ}$ and $b\le$-20$^{\circ}$ 2MASS 2-point angular correlation functions
at large scales for example. It has also recently been suggested that the southern galactic cap may contain a large local under-density in the galaxy 
distribution covering \gsim4000~deg$^2$ around the Southern Galactic Pole to $r\approx$300\mpc, which may be at odds with the form of the $\Lambda$CDM 
$P(k)$ at large scales \citep{fri4,fri3,fri,bus}. It is therefore interesting to compare the form of high-order clustering statistics in the 
galactic caps. 

We plot the 2-point angular correlation functions and high-order scaling parameters for $b\ge$10$^{\circ}$ and $b\le$-10$^{\circ}$ $K_s<$13.5 2MASS 
galaxies in the third column of Fig.~\ref{fig:wp2}. We observe a similar discrepancy in the 2-point function as determined previously by \citet{mal} 
with a steeper slope in the southern $\bar{\omega}_2$ at large scales. However, the high-order scaling parameters in the northern and southern 
local galaxy distributions, for $p\le$4 at least, are consistent with each other and constant values over $\approx$3 orders of magnitude of 
angular scales to $\theta$\lsim20$^{\circ}$. At higher orders there are differences between the two which increase with $p$. Whether these differences 
are due to real north-south differences in the local galaxy distribution, or simply due to other systematic effects is unclear.

\section {Discussion}

In the previous section, high-order angular and real space hierarchical scaling parameters were determined to extremely large scales ($r$\lsim
40\mpc) from a sample of 650$\,$745 $K_s<$13.5 2MASS galaxies. Unlike previous high-order clustering analyses, the scaling relations for $p$\lsim 4 
are relatively robust even when the most prominent clusters are removed from the sample.

As described in sections 1 and 2, high-order clustering statistics have significant implications for the nature of the primordial density distribution 
and also the way in which galaxies trace the underlying mass distribution. In the following section we examine the consistency of our results 
with Gaussian initial conditions and hierarchical clustering and discuss the implications for non-Gaussian models of the primordial density 
distribution. Assuming the former, we also infer constraints on non-linear galaxy bias.

\subsection{Implications for the primordial density field}

\begin{figure}
\begin{center}
\centerline{\epsfxsize = 3.5in
\epsfbox{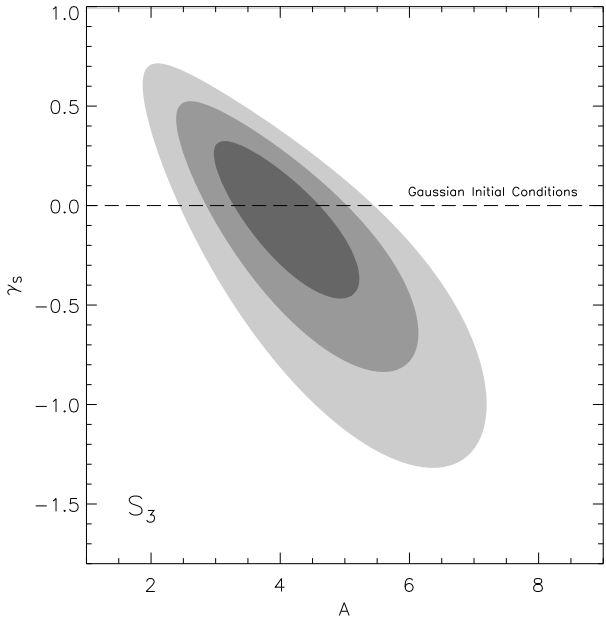}}
\caption{Filled contours representing the 1$\sigma$, 2$\sigma$ and 3$\sigma$ confidence regions for the real space galaxy skewness
slope and amplitude determined from $\chi^2$ fits (accounting for the covariance) to the 2MASS $|b|\ge$10$^{\circ}$ $K_s<$13.5 $S_3$ parameter
(as shown in Fig.~\ref{fig:wp}) at large scales (4.0$<\bar{r}<$40\mpc ). The best fit parameters are $A$=4.0 and $\gamma_S$=-0.02, where we model the
skewness using $S_3=A~r^{\gamma_S}$. The dashed line indicates the predicted constant form (i.e. $\gamma_S$=0) for $S_3$ in the case of a Gaussian
distribution of primordial density fluctuations. }
\label{fig:gamma}
\end{center}
\end{figure}

\begin{figure}
\begin{center}
\centerline{\epsfxsize = 3.5in
\epsfbox{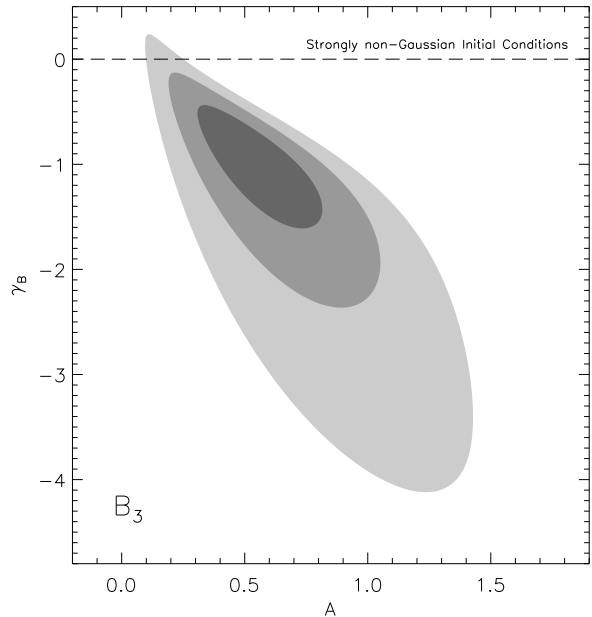}}
\caption{Filled contours representing the 1$\sigma$, 2$\sigma$ and 3$\sigma$ confidence regions for the real space $B_3$ parameter (see
equation~\ref{equation:sngic}) slope and amplitude determined from $\chi^2$ fits as in Fig.~\ref{fig:gamma}. The best fit parameters are $A$=0.53 and
$\gamma_B$=-0.93, where we use $B_3=A~r^{\gamma_B}$ as before. The dashed line indicates the predicted constant form for $B_3$ in the
case of a strongly non-Gaussian distribution of primordial density fluctuations.}
\label{fig:gamma2}
\end{center}
\end{figure}

\subsubsection{Consistency with Gaussian initial conditions}

The nature of the primordial distribution of density fluctuations is predicted to be close to or exactly Gaussian in standard inflationary models 
\citep{fal,gan,les,wan,mald,acq}. This occurs as a consequence of the slow-roll conditions on the inflation potential which require the potential 
energy of the field to dominate over the kinetic energy in order to produce a phase of accelerated expansion which lasts for a sufficiently long 
period of time. Deviations from Gaussianity under these assumptions depend on the inflationary model but are generally extremely small; in recent 
power spectrum analyses for example it is assumed that the primordial density field has exactly random phases \citep[e.g.][]{col3,fri2}.

Under the assumption of Gaussian initial conditions, a hierarchical scaling (see equation~\ref{equation:3dhier}) of the 
high-order moments of the dark matter and galaxy density fields \citep{fry2} is expected through the evolution under gravitational instability of 
the primordial density fluctuations \citep[e.g.][]{peb4,fry,bou2,ber,ber2,gaz3,bau2}. This holds on scales where the gravitational collapse of dark 
matter haloes evolves linearly; it is important therefore to examine high-order moments in the linear and quasi-linear regime.

It is clear from Fig.~\ref{fig:wp} that our results provide an excellent agreement with the expected scaling relations at large scales, with 
$s_p$ and $S_p$ parameters consistent with constant values over several orders of magnitude in scale for $p\le$7. As noted previously by the 2dFGRS 
\citep{bau2}, this hierarchical scaling extends to smaller scales than expected by perturbation theory predictions \citep{ber3}. This consistency 
with Gaussian initial conditions is also apparent from Tables~\ref{table:sptab} and \ref{table:sptab2} where we perform fits to the scaling 
parameters to unprecedented scales; there is good agreement between the small and large scale fits (although there is a small correlation between the 
two). Alternatively, we can examine the measured slope of the $S_p$ parameter at the scales of interest; in Fig.~\ref{fig:gamma} we determine 
confidence limits for the slope and amplitude of the $S_3$ parameter at large scales (4.0$<\bar{r}<$40\mpc ) as shown in Fig.~\ref{fig:wp}. The 
constraints provide excellent agreement with the expected constant value as a function of scale, with a best fit slope of 
$\gamma_S=-0.02^{+0.34}_{-0.44}$ (where $S_3\propto r^{\gamma_S}$, marginalising over the normalisation).

This consistency with hierarchical scaling represents a departure from recent constraints on high-order correlation functions. The APM 
\citep{gaz} and Edinburgh-Durham Southern Galaxy Catalogue \citep{sza} surveys observe significant upturns in the angular skewness at scales of 
$\theta$\gsim1$^{\circ}$ (corresponding to $r$\gsim4\mpc). Also, analysis of the 2dFGRS \citep{cro,bau2} indicates rising $S_p$ parameters. The 
issue is complicated by the fact that these measurements are not independent as there is significant overlap between the observed survey fields. 
Coupled with this, the 2dFGRS identify two superclusters which significantly alter their results; when removed there is good agreement with the 
expected hierarchical scaling. However, due to the survey volume the scaling parameters are determined only below $r\approx$10\mpc~ and do not probe 
the larger scales of interest here.

\subsubsection{Constraints on non-Gaussianity}

While the assumption of Gaussian initial conditions is acceptable assuming the validity of the simplest inflationary models, there also exist 
alternative models of inflation in which deviations from a Gaussian form for the primordial density field are predicted.
For example, it is possible to introduce non-Gaussianity if the scalar field driving inflation has more than one component  
\citep[e.g.][]{kof,ber5}; strongly non-Gaussian initial conditions are also a feature of models in which the inflaton is not a slowly 
rolling scalar field but a fast moving ghost condensate \citep{ark}. Additionally, it is possible to construct models in which the initial conditions 
are non-Gaussian as a result of non-linear structures, such as cosmic strings or global textures, within the primordial density field 
\citep[e.g.][]{gaz5,ave,gaz4}.

When considering the effect of non-Gaussian initial conditions on moments of the local galaxy density field, it is convenient to consider the 
departure from primordial Gaussianity in two ways. Here, we consider the strongly non-Gaussian regime as might arise from cosmic strings or 
textures; weak departures from Gaussianity as might arise from the various models of inflation described above are more difficult to constrain since 
the effect can be a shift in the $s_p$ and $S_p$ parameter amplitudes rather than a change in slope. In the case of strong departures from Gaussian 
initial conditions, the scaling of high-order moments of the density field is expected:

\begin{equation}
\bar{\xi}_{p} = B_{p}~ \bar{\xi}_{2}^{p/2} ,
\label{equation:sngic}
\end{equation}

\noindent where for non-Gaussian models seeded by topological defects the $B_p$ scaling parameters (not to be confused with $B_p$ used 
in equation~\ref{equation:3dpt}) are expected to be constant at large scales and of order unity \citep{tur,gaz5,ber3}. Equivalently, the typical 
signature of strong non-Gaussianity in the $S_p$ scaling parameters is a slope of $\gamma_S$\gsim0.6 (where $S_p\propto r^{\gamma_S}$) at large 
scales, with a characteristic minimum at $r\approx$10\mpc~ \citep{gaz5}.

In Fig.~\ref{fig:gamma2} we constrain the slope and amplitude of the $B_3$ parameter through $\chi^2$ fits to the 
2MASS results (measuring the projected $B_3=\bar{\omega}_3/\bar{\omega}_2^{3/2}$ and using the transformation to real space described in 
equation~\ref{equation:3dpt}) at large scales (4.0$<\bar{r}<$40\mpc ). We are able to reject the scaling expected in strongly non-Gaussian 
models (i.e. constant $B_p$ parameters, $\gamma_B$=0) as described in equation~\ref{equation:sngic}, and therefore non-Gaussian models seeded by 
topological defects, at the $\approx$2.5$\sigma$ confidence level.

\subsection{Non-linear galaxy bias}

High-order clustering analysis also represents a powerful probe of the way in which galaxies trace the underlying mass distribution. We have 
previously noted that the galaxy bias associated with the variance of the density field has been constrained to $b_1>1$ in the 
$K_s$-band at the $>3\sigma$ level \citep{fri2} and $b_1\approx 1$ for optically-selected galaxies. It is unclear whether deviations from the 
linear bias model exist, and if they do, how this coincides with current theories of galaxy evolution.

Assuming Gaussian initial conditions, we use predictions from perturbation theory for the dark matter skewness and the 
relationship between the dark matter and galaxy skewness (equations~\ref{equation:skew} and \ref{equation:2dpt}) to compute the $K_s$-band non-linear 
bias coefficient $c_2=b_2/b_1$ implied by our results. We use an $n$=-2 power spectrum slope \citep{per,col3} and a $K_s$-band linear bias of 
$b_1$=1.39$\pm$0.12 measured from the $K_s<$13.5 2MASS galaxy angular power spectrum \citep{fri2}. Since these predictions from perturbation theory 
are valid only in the quasi-linear and linear regimes \citep{ber4}, we use only the fits on large scales. First, we use the best fit galaxy skewness 
in the range 4.0$<\bar{r}<$40\mpc~(see Table~\ref{table:sptab2}); we find that c$_{2}$=0.57$\pm$0.41. Since the conversion from $s_3$ to $S_3$ 
becomes increasingly uncertain at scales of $\theta>2^{\circ}$ \citep{gaz}, we also use a narrower range of 4.0$<\bar{r}<$10\mpc~(equivalent to 
1$^{\circ}.0<\theta<2^{\circ}$.5). Using the best fit galaxy skewness in this range of $S_3$=4.01$\pm$0.34, we find that $c_2=0.57\pm0.33$.

We therefore detect a {\em positive} quadratic contribution to the $K_s$-band galaxy bias at the $\approx 2\sigma$ level. This means that the density 
of 2MASS galaxies rises more quickly than the mass density contrast. This differs from all previous constraints on the $c_2$ non-linear bias 
parameter which have been {\em negative}, most recently with constraints from the 2dFGRS which limit $b_1=0.94^{+0.13}_{-0.11}$ and 
$c_2=-0.36^{+0.13}_{-0.09}$ in the optical $b_J$-band \citep{gaz7}, and also from the IRAS PSCz catalogue of $b_1=0.83\pm0.13$ and $c_2=-0.50\pm0.48$ 
in the infra-red \citep{fel2}. This compares to near infra-red $K_s$-band constraints of $b_1=1.39\pm0.12$ \citep{fri2} and the optimal constraint in 
this work of $c_2=0.57\pm0.33$.

It is possible to understand these results by examining the analytic predictions of \citet{mo} for the high-order bias coefficients, formed via 
the \citet{pre} formalism and an initially Gaussian density field. From this the first two terms in the Taylor expansion of   
equation~\ref{equation:taylor} are predicted to be:

\begin{equation}
b_1=1+\frac{\nu^2-1}{\delta_c}
\label{equation:b1}
\end{equation}

\begin{equation}
b_2=2 \left( 1-\frac{17}{21} \right) \frac{\nu^2-1}{\delta_c} + \left(\frac{\nu}{\delta_c}\right)^2(\nu^2-3) ,
\label{equation:b2}
\end{equation}

\noindent where $\nu=\delta_c/\sigma(M)$, $\sigma(M)$ denotes the linear {\it rms} fluctuation on the mass scale of a dark matter halo of 
mass $M$ and $\delta_c$ is the linear theory over-density at the time of collapse (for reference $\delta_c$=1.686 for $\Omega$=1). 

This relatively simplistic scenario is not able to provide accurate quantitative predictions which match the observational results above. However, by 
considering haloes of differing mass it is possible, qualitatively at least, to understand these apparently contradictory constraints on $c_2$. For 
instance, if we consider the most massive haloes for which $\nu^2$\gsim 3, \citet{mo} predict $b_1>$1 and $c_2>$0. In contrast, for typical mass 
haloes for which $\nu^2\approx$1, bias parameters of $b_1\approx$1 and $c_2<$0 might be expected. With this in mind, it is possible to understand why 
a near infra-red survey, which is more sensitive to early-type galaxies than optical or infra-red surveys \citep[e.g.][~respectively]{jar2,cro2,oli}, 
produces galaxy samples with higher values for the linear and quadratic bias parameters.

\section {Conclusions}

We have measured reduced angular correlation functions, $\bar{\omega}_p$, to ninth order using 650~745 galaxies selected from the 2MASS 
extended source catalogue. From our estimates for the angular correlation functions, we have determined the projected and real space 
hierarchical scaling parameters, $s_p$ and $S_p$ respectively. The prime motivation for such analysis is to test the hierarchical scaling hypothesis 
which predicts these parameters to be constant in the linear and quasi-linear regimes \citep[e.g.][]{peb4,fry,bou2,ber}. As such we are able to probe 
the primordial density field and constrain various models of inflation and structure formation. High-order clustering analysis also allows us to probe 
the way in which galaxies trace the underlying mass distribution; a negative offset between $\Lambda$CDM predictions and observations by 
the 2dFGRS have recently been interpreted as evidence for a quadratic contribution to the galaxy bias, although these conclusions are based on 
constraints in the weakly non-linear regime \citep{gaz7}.

The most comparable recent work are the analyses of the 2dFGRS \citep{cro,bau2} and APM surveys \citep{gaz}; 
the galaxy sample used in this work represents an order of magnitude increase in volume and solid angle over each respectively. Previous analyses of 
high-order clustering statistics have proved extremely challenging; due to the relatively small volumes probed in 3-dimensional analyses for instance, 
the results have been sensitive to the presence of clusters and superclusters within the galaxy sample. Not only this, but direct comparisons with 
perturbation theory have also proved difficult since the statistical uncertainty at large scales is considerable, and frequently the covariance in 
the statistics has been ignored leading to unrealistically small errors. In addition, the results for the hierarchical scaling parameters have 
consistently displayed a puzzling upturn in values at $r\approx$4\mpc , consistent with some models of structure formation with strongly non-Gaussian 
initial conditions \citep{gaz5,gaz4,whi,ber3}. 

Here, we are able to determine the scaling parameters to high accuracy to unprecedented scales, $r$\lsim 100\mpc . We also carry out a full covariance 
analysis in order to take account of correlations in the datapoints at different cell radii. We are therefore in a position to make direct comparisons 
with the predictions of perturbation theory since we probe well into the quasi-linear and linear regimes and we have a good understanding of the 
statistical uncertainty. However, since we are working with a projected galaxy sample, we have to convert the associated angular scaling parameters 
to real space via a transformation which becomes increasingly uncertain on large scales ($\theta>$2$^{\circ}$). Nevertheless, this work currently 
provides the best estimates of high-order clustering statistics at large scales. We are able to reach a number of conclusions:

(i) Our results are in line with the expected hierarchical scaling relation, with $s_p$ and $S_p$ parameters consistent with constant values over 
$\approx$3 orders of magnitude in scale to $r\approx$40\mpc~for $p\le$7; we constrain the slope of $S_3$ to $\gamma_S=-0.02^{+0.34}_{-0.44}$ (where 
$S_p\propto r^{\gamma}$). Such a scaling pattern is expected if an initially Gaussian density field evolves under the action of gravitational 
instability \citep[e.g.][]{peb4,fry,bou2,ber}. This result is in contrast to recent results drawn from the 2dFGRS, APM and EDSGC surveys 
\citep{gaz,sza,bau2,cro} which display rising scaling parameters at large scales. 

(ii) The scaling parameters are relatively robust to the removal of the largest over-density, the Shapley supercluster, although the best fit slope 
of $s_3$ at large scales (1$^{\circ}.0<\theta<$10$^{\circ}$) becomes steeper, yet is still consistent with a constant value 
($\gamma=-0.27^{+0.25}_{-0.29}$). We also use a more stringent cut by removing the ten most 
over-dense regions (see Fig.~\ref{fig:sky} and Table~\ref{table:clusters}) corresponding to a cut to the main sample of 6.3\% of the galaxies and 
2.6\% of the total solid angle; the resulting $s_3$ parameter remains broadly consistent with the result for the main sample, although with a 
slightly steeper best fit slope at large scales ($\gamma=-0.33^{+0.26}_{-0.29}$).

(iii) Since strong non-Gaussianity in the primordial density field, as might be expected in models seeded by topological defects such as cosmic 
strings or global textures \citep{ave,gaz4,gaz5}, results in a strong upturn in the hierarchical scaling parameters at large scales 
\citep[e.g.][]{ber3}, we are able to reject strongly non-Gaussian initial conditions, producing relations of the form
$\bar{\xi}_p\propto\bar{\xi}_2^{p/2}$, at the $\approx$2.5$\sigma$ confidence level. 

(iv) We compare our constraints on $S_3$ at large scales (where we have assumed a constant value) to predictions from perturbation theory. We detect a 
significant deviation consistent with a non-linear, quadratic contribution to the $K_s$-band galaxy bias, parameterised as $c_2$=0.57$\pm$0.33 
(derived from fits in the range 4.0$<\bar{r}<$10\mpc ), implying that the 2MASS galaxy density rises more quickly than the mass density contrast. 
This {\em positive} result represents a significant difference from the negative values found previously; constraints on $c_2$ from the 
optically-selected 2dFGRS and the infra-red PSCz samples yield negative values. We explain these apparently 
contradictory results through an examination of the model of \citet{mo}, which predicts bias parameters of $b_1>1$ (the linear bias) and $c_2>$0 (the 
quadratic bias) if the surveyed galaxies typically reside in large mass haloes. Similarly, as observed previously, we might expect $b_1\approx$1 and 
$c_2<$0 from galaxy samples in which late-type galaxies are over-represented.

\section*{Acknowledgements} 

This publication makes use of data products from the 2 Micron All-Sky Survey, which is a joint project of the University of
Massachusetts and the Infrared Processing and Analysis Centre/California Institute of Technology, funded by the Aeronautics and
Space Administration and the National Science Foundation. We thank Carlton Baugh for kindly reading through this paper and providing useful feedback. 
We also thank Enrique Gazta\~{n}aga for useful discussion.

\label{lastpage}


\vspace*{-0.1cm}

\begin{thebibliography}{99}

\bibitem[\protect\citeauthoryear{Acquaviva et al.}{2003}]{acq} Acquaviva, V., Bartolo, N., Matarrese, S. \& Riotto, A. 2003, Nucl. Phys. B, 667, 119
\bibitem[\protect\citeauthoryear{Arkani-Hamed et al.}{2004}]{ark} Arkani-Hamed, N., Creminelli, P., Mukohyama, S. \& Zaldarriaga, M. 2004, J. 
Cosmol. Astropart. Phys., 04, 001 
\bibitem[\protect\citeauthoryear{Avelino et al.}{1998}]{ave} Avelino, P.P., Shellard, E.P., Wu, J.H. \& Allen, B. 1998, ApJ, 507, L101
\bibitem[\protect\citeauthoryear{Baugh \& Efstathiou}{1993}]{bau} Baugh, C.M. \& Efstathiou, G. 1993, MNRAS, 265, 145
\bibitem[\protect\citeauthoryear{Baugh et al.}{1995}]{bau3} Baugh, C.M., Gazta\~{n}aga, E. \& Efstathiou, G. 1995, MNRAS, 274, 1049
\bibitem[\protect\citeauthoryear{Baugh}{1996}]{bau4} Baugh, C.M. 1996, MNRAS, 280, 267
\bibitem[\protect\citeauthoryear{Baugh et al.}{2004}]{bau2} Baugh, C.M. et al. 2004, MNRAS, 351, L44
\bibitem[\protect\citeauthoryear{Bernardeau}{1992}]{ber} Bernardeau, F. 1992, ApJ, 392, 1
\bibitem[\protect\citeauthoryear{Bernardeau et al.}{2002}]{ber3} Bernardeau, F., Colombi, S., Gazta\~{n}aga, E. \& Scoccimarro, R. 2002, Phys.
Rep., 367, 1
\bibitem[\protect\citeauthoryear{Bernardeau}{1994a}]{ber2} Bernardeau, F. 1994a, A\&A, 291, 697
\bibitem[\protect\citeauthoryear{Bernardeau}{1994b}]{ber4} Bernardeau, F. 1994b, ApJ, 433, 1
\bibitem[\protect\citeauthoryear{Bernardeau \& Uzan}{2003}]{ber5} Bernardeau, F. \& Uzan, J.P. 2003, Phys. Rev. D., 67, 121301
\bibitem[\protect\citeauthoryear{Bouchet et al.}{1992}]{bou2} Bouchet, F.R., Juszkiewicz, R., Colombi, S. \& Pellat, R.
1992, ApJ, 394, L15
\bibitem[\protect\citeauthoryear{Bouchet et al.}{1993}]{bou} Bouchet, F.R., Strauss, M., Davis, M., Fisher, K., Yahil, A. \& Huchra, J. 
1993, ApJ, 417, 36
\bibitem[\protect\citeauthoryear{Busswell et al.}{2004}]{bus} Busswell, G.S., Shanks, T., Outram, P.J., Frith, W.J.,
Metcalfe, N. \& Fong, R. 2004, MNRAS, 354, 991
\bibitem[\protect\citeauthoryear{Cabella et al.}{2005}]{cab} Cabella, P., Liguori, M., Hansen, F.K., Marinucci, D.,
Matarrese, S., Moscardini, L. \& Vittorio, N. 2005, MNRAS, 358, 684
\bibitem[\protect\citeauthoryear{Carlberg et al.}{2000}]{car} Carlberg, R.G., Yee, H.K., Morris, S.L., Lin, H., Hall, P.B., Patton, D., Sarwicki, M. 
\& Shepherd, C.W. 2000, ApJ, 542, 57
\bibitem[\protect\citeauthoryear{Cole et al.}{2001}]{cole} Cole, S.M. et al. 2001, MNRAS, 326, 555
\bibitem[\protect\citeauthoryear{Cole et al.}{2005}]{col3} Cole, S.M. et al. 2005, accepted by MNRAS, astro-ph/0501174
\bibitem[\protect\citeauthoryear{Croton et al.}{2004}]{cro} Croton, D.J. et al. 2004, MNRAS, 352, 1232
\bibitem[\protect\citeauthoryear{Croton et al.}{2005}]{cro2} Croton, D.J. et al. 2005, MNRAS, 356, 1155
\bibitem[\protect\citeauthoryear{Falk et al.}{1993}]{fal} Falk, T., Rangarajan, R. \& Srednicki, M. 1993, ApJ, 403, L1
\bibitem[\protect\citeauthoryear{Feldman et al.}{2001}]{fel2} Feldman, H.A., Frieman, J.A., Fry, J.N. \& Scoccimarro, R. 2001, Phys. Rev. Lett., 
86, 1434
\bibitem[\protect\citeauthoryear{Frith et al.}{2003}]{fri} Frith, W.J., Busswell, G.S., Fong, R., Metcalfe, N. 
\& Shanks, T. 2003, MNRAS, 345, 1049
\bibitem[\protect\citeauthoryear{Frith et al.}{2004}]{fri3} Frith, W.J., Outram, P.J. \& Shanks, T. 2004, ASP Conf. Proc., Volume 329, 49
\bibitem[\protect\citeauthoryear{Frith et al.}{2005a}]{fri4} Frith, W.J., Shanks, T. \& Outram, P.J. 2005a, MNRAS, 361, 701
\bibitem[\protect\citeauthoryear{Frith et al.}{2005b}]{fri2} Frith, W.J., Outram, P.J. \& Shanks, T. 2005b, submitted to MNRAS, astro-ph/0507215
\bibitem[\protect\citeauthoryear{Fry}{1984}]{fry} Fry, J.N. 1984, ApJ, 279, 499
\bibitem[\protect\citeauthoryear{Fry \& Gazta\~{n}aga}{1993}]{fry2} Fry, J.N. \& Gazta\~{n}aga, E. 1993, ApJ, 413, 447
\bibitem[\protect\citeauthoryear{Gangui et al.}{1994}]{gan} Gangui, A., Lucchin, F., Matarrese, S. \& Mollerach, S. 1994, ApJ, 430, 447
\bibitem[\protect\citeauthoryear{Gazta\~{n}aga}{1994}]{gaz} Gazta\~{n}aga, E. 1994, MNRAS, 268, 913
\bibitem[\protect\citeauthoryear{Gazta\~{n}aga \& Frieman}{1994}]{gaz6} Gazta\~{n}aga, E. \& Frieman, J.A. 1994, ApJ, 437, L13
\bibitem[\protect\citeauthoryear{Gazta\~{n}aga \& Baugh}{1995}]{gaz3} Gazta\~{n}aga, E. \& Baugh, C.M. 1995, MNRAS, 273, L1
\bibitem[\protect\citeauthoryear{Gazta\~{n}aga \& Mahonen}{1996}]{gaz5} Gazta\~{n}aga, E. \& Mahonen, P. 1996, ApJ, 462, L1 
\bibitem[\protect\citeauthoryear{Gazta\~{n}aga \& Fosalba}{1998}]{gaz4} Gazta\~{n}aga, E. \& Fosalba, P. 1998, MNRAS, 301, 524
\bibitem[\protect\citeauthoryear{Gazta\~{n}aga \& Bernardeau}{1998}]{gaz2} Gazta\~{n}aga, E. \& Bernardeau, F. 1998, A\&A, 331, 829 
\bibitem[\protect\citeauthoryear{Gazta\~{n}aga et al.}{2005}]{gaz7} Gazta\~{n}aga, E., Norberg, P., Baugh, C.M. \& Croton, D.J. 2005, 
submitted to MNRAS, astro-ph/0506249
\bibitem[\protect\citeauthoryear{Groth \& Peebles}{1977}]{gro} Groth, E.J. \& Peebles, P.J.E. 1977, ApJ, 217, 385
\bibitem[\protect\citeauthoryear{Hamilton et al.}{1991}]{ham} Hamilton, A.J., Kumar, P., Lu, E. \& Matthews, A. 1991, ApJ, 374, 1
\bibitem[\protect\citeauthoryear{Hawkins et al.}{2003}]{haw} Hawkins, E. et al. 2003, MNRAS, 346, 78
\bibitem[\protect\citeauthoryear{Hoyle et al.}{2000}]{hoy} Hoyle, F., Szapudi, I. \& Baugh, C.M. 2000, MNRAS, 317, 51
\bibitem[\protect\citeauthoryear{Jarrett et al.}{2000}]{jar} Jarrett, T.H., Chester, T., Cutri, R., Schneider, S.,
Skrutskie, M. \& Huchra, J.P. 2000, AJ, 119, 2498
\bibitem[\protect\citeauthoryear{Jarrett}{2004}]{jar2} Jarrett, T.H. 2004, astro-ph/0405069
\bibitem[\protect\citeauthoryear{Jenkins et al.}{1998}]{jen} Jenkins, A. et al. 1998, ApJ, 499, 20
\bibitem[\protect\citeauthoryear{Jones et al.}{2004}]{jon} Jones, D.H. et al. 2004, MNRAS, 355, 747
\bibitem[\protect\citeauthoryear{Juszkiewicz et al.}{1993}]{jus} Juszkiewicz, R., Bouchet, F.R. \& Colombi, S. 1993, ApJ, 412, 9
\bibitem[\protect\citeauthoryear{Kofman \& Pogosyan}{1988}]{kof} Kofman, L. \& Pogosyan, D.Y. 1988, Phys. Lett. B, 214, 508
\bibitem[\protect\citeauthoryear{Lesgourgues}{1997}]{les} Lesgourgues, J., Polarski, D. \& Starobinsky, A.A. 1997, Nucl. Phys. B, 497, 479
\bibitem[\protect\citeauthoryear{Loveday}{2000}]{lov} Loveday, J. 2000, MNRAS, 312, 517
\bibitem[\protect\citeauthoryear{Maldacena}{2002}]{mald}Maldacena, J. 2002, JHEP, 0305, 013
\bibitem[\protect\citeauthoryear{Maller et al.}{2005}]{mal}Maller, A.H., McIntosh, D.H., Katz, N. \& Weinberg, M.D. 2005, ApJ, 619, 147
\bibitem[\protect\citeauthoryear{Maller et al.}{2003}]{mal2}Maller, A.H., McIntosh, D.H., Katz, N. \&
Weinberg, M.D. 2003, ApJ, 598, 1
\bibitem[\protect\citeauthoryear{Mo et al.}{1997}]{mo} Mo, H.J., Jing, Y.P. \& White, S.D. 1997, MNRAS, 284, 189
\bibitem[\protect\citeauthoryear{Oliver et al.}{1996}]{oli} Oliver, S.J.  et al. 1996, MNRAS, 280, 673
\bibitem[\protect\citeauthoryear{Pan \& Szapudi}{2005}]{pan} Pan, J. \& Szapudi, I. 2005, submitted to MNRAS, astro-ph/0505422
\bibitem[\protect\citeauthoryear{Peacock \& Dodds}{1994}]{pea} Peacock, J.A. \& Dodds, S.J. 1994, MNRAS, 267, 1020
\bibitem[\protect\citeauthoryear{Peebles}{1980}]{peb4} Peebles, P.J.E. 1980, Principles of Physical Cosmology, Princeton University Press 
\bibitem[\protect\citeauthoryear{Percival et al.}{2001}]{per} Percival, W.J. et al. 2001, MNRAS, 327, 1297
\bibitem[\protect\citeauthoryear{Press \& Schechter}{1974}]{pre} Press, W.H. \& Schechter, P. 1974, ApJ, 187, 425
\bibitem[\protect\citeauthoryear{Salopek \& Bond}{1991}]{sal} Salopek, D.S. \& Bond, J.R. 1991, Phys. Rev. D, 43, 1005
\bibitem[\protect\citeauthoryear{Saunders et al.}{1991}]{sau} Saunders, W. et al. 1991, Nat, 349, 32
\bibitem[\protect\citeauthoryear{Schlegel et al.}{1998}]{sch} Schlegel, D.J., Finkbeiner, D.P. \& Davis, M. 1998, ApJ, 500, 525
\bibitem[\protect\citeauthoryear{Scoccimarro et al.}{2003}]{sco} Scoccimarro, R., Sefusatti, E. \& Zaldarriaga, M. 2003, Phys. Rev. D, 69, 103513
\bibitem[\protect\citeauthoryear{Silk \& Juszkiewicz}{1991}]{sil} Silk, J. \& Juszkiewicz, R. 1991, Nat, 353, 386
\bibitem[\protect\citeauthoryear{Spergel et al.}{2003}]{spe} Spergel, D.N. et al. 2003, ApJS, 148, 175
\bibitem[\protect\citeauthoryear{Szapudi \& Gazta\~{n}aga}{1998}]{sza} Szapudi, I. \& Gazta\~{n}aga, E. 1998, MNRAS, 300, 493
\bibitem[\protect\citeauthoryear{Szapudi et al.}{2000}]{sza2} Szapudi, I., Colombi, S., Jenkins, A. \& Colberg, J. 2000, MNRAS, 313, 725
\bibitem[\protect\citeauthoryear{Turok \& Spergel}{1991}]{tur} Turok, N. \& Spergel, D.N. 1991, Phys. Rev. Lett., 66, 3093
\bibitem[\protect\citeauthoryear{Verde et al.}{2002}]{ver} Verde, L. et al. 2002, MNRAS, 335, 432
\bibitem[\protect\citeauthoryear{Wang \& Kamionkowski}{2000}]{wan} Wang, L. \& Kamionkowski, M. 2000, Phys. Rev. D, 61, 063504
\bibitem[\protect\citeauthoryear{Weinberg \& Cole}{1992}]{wei} Weinberg, D.H. \& Cole, S.M. 1992, MNRAS, 259, 652
\bibitem[\protect\citeauthoryear{Wilson}{2003}]{wil} Wilson, G. 2003, ApJ, 585, 191
\bibitem[\protect\citeauthoryear{White}{1999}]{whi} White, M. 1999, MNRAS, 310, 511
\bibitem[\protect\citeauthoryear{Zehavi et al.}{2004}]{zeh2} Zehavi, I. et al. 2004, ApJ, 608, 16

\end{thebibliography}
\end{document}